# Quantum mechanics emerging from stochastic dynamics of virtual particles

Roumen Tsekov

Department of Physical Chemistry, University of Sofia, 1164 Sofia, Bulgaria

It is demonstrated how quantum mechanics emerges from the stochastic dynamics of force-carriers. It is shown that the quantum Moyal equation corresponds to some dynamic correlations between the momentum of a real particle and the position of a virtual particle, which are not present in classical mechanics. The new concept throws light on the physical meaning of quantum theory, showing that the Planck constant square is a second-second cross-cumulant.

According to modern physics, the Newtonian interactions in classical mechanics occur via exchange of virtual particles [1]. These force-carriers transmit the long-range interactions among the real particles and interact only locally with the latter. Generally, it is expected that the virtual particle dynamics is stochastic, which will obviously result in random Newtonian potentials. Therefore, the stochastic motion of the virtual particles can cause the quantum dynamics [2]. For instance, the stochastic electrodynamics is an important example, which is already proposed for the origin of quantum mechanics [3]. In the present paper it is shown, how quantum mechanics emerges from the stochastic dynamics of force-carriers. It is also demonstrated that the quantum Moyal equation [4-6] corresponds to some statistical correlations between the momentum of a real particle and the position of the virtual particles, being not present in classical mechanics. The new paradigm throws light on the physical meaning of the Planck constant, which is the quintessence of quantum theory.

For the sake of transparency, let us consider first a real and a virtual particles, located at positions $r$ and $R$, respectively. Since by definition there is no long-range interactions between the real particle and force-carriers, their interaction potential $\varepsilon\delta(r-R)$ is a Dirac delta-function with a specific interaction parameter $\varepsilon$. The latter accounts for the type of the interaction, e.g. electrostatic, gravitational, etc. In general, the position of the virtual particle is expected to be stochastic and, hence, the interaction delta-potential above is a random one. The corresponding force, acting on the real particle, reads $-\varepsilon\partial_r\delta(r-R)$. Introducing the phase-space distribution density $W(p,r,t)$ of the real particle, one can write its dynamic balance in the form

$$\partial_t W + p \cdot \partial_r W / m = \partial_p \cdot \int_{-\infty}^{\infty} \varepsilon \partial_r \delta(r-R) F(R,p,r,t) dR = \partial_p \cdot (\varepsilon \partial_R F)_{R=r} \tag{1}$$

The interaction between the real and virtual particles is expressed in Eq. (1) via a collision integral, where $F(R,p,r,t)$ is the probability density for the virtual particle to occupy the position $R$ and for the real particle to have momentum $p$ and coordinate $r$ at time $t$. The real particle probability density $W$ can be derived from $F$ via a simple integration over $R$

$$W(p,r,t) \equiv \int_{-\infty}^{\infty} F(R,p,r,t) dR \qquad \rho(R,t) \equiv \int_{-\infty}^{\infty} \int_{-\infty}^{\infty} F(R,p,r,t) dp dr \tag{2}$$

while $\rho(R,t)$ is the local density of the virtual particles. A macro-particle interacts with many force carriers at once and thus stochastic forces cancel each other. Hence, in classical mechanics the force-carriers distribution is independent of the real particle and the joint probability density factorizes in a product of the virtual particle density and probability density of the real particle

$$F(R,p,r,t) = \rho(R,t) W(p,r,t) \tag{3}$$

In this case Eq. (1) reduces straightforward to the Liouville equation from classical mechanics

$$\partial_t W + p \cdot \partial_r W / m = \partial_r U \cdot \partial_p W \qquad U(r,t) \equiv \int_{-\infty}^{\infty} \varepsilon \delta(r-R) \rho(R,t) dR = \varepsilon \rho(r,t) \tag{4}$$

The corresponding Newtonian potential $U$ follows exactly the distribution of the virtual particles in the system, which is not perturbed by the motion of the real particle.

There are indications in quantum mechanics, however, that the motion of the real particle affects the distribution of virtual particles in the system. The Liouville equation (4) changes to the Wigner-Liouville or Moyal equation bellow

$$\partial_t W + p \cdot \partial_r W / m = \sum_{n=0}^{\infty} \frac{(i\hbar/2)^{2n}}{(2n+1)!} \partial_r^{2n+1} U \cdot \partial_p^{2n+1} W \qquad (5)$$

Keeping in mind the definition of the Newtonian potential $U = \varepsilon\rho$, the juxtaposition of the Wigner-Liouville equation (5) and Eq. (1) unveils a possible expression for the joint distribution density

$$F(R,p,r,t) = \sum_{n=0}^{\infty} \frac{(i\hbar/2)^{2n}}{(2n+1)!} \partial_R^{2n} \rho \cdot \partial_p^{2n} W = \rho W + \sum_{n=1}^{\infty} \frac{(i\hbar/2)^{2n}}{(2n+1)!} \partial_R^{2n} \rho \cdot \partial_p^{2n} W \qquad (6)$$

As is seen, the last term represents a correlation function between the virtual and real particles, which vanishes in the classical limit at $\hbar \to 0$. The analysis of the probability density (6) is easier in the Fourier space, where the corresponding characteristic function is given by

$$\tilde{F}(K,q,k,t) \equiv \int_{-\infty}^{\infty}\int_{-\infty}^{\infty}\int_{-\infty}^{\infty} \exp(iK \cdot R + iq \cdot p + ik \cdot r) F(R,p,r,t) dR dp dr \qquad (7)$$

Substituting here Eq. (6) yields the characteristic function

$$\tilde{F}(K,q,k,t) = \tilde{\rho}(K,t) \frac{\sin(\hbar K \cdot q/2)}{\hbar K \cdot q/2} \tilde{W}(q,k,t) \qquad (8)$$

being expressed by the characteristic functions of the marginal distributions from Eq. (2). Taking a logarithm from Eq. (8) provides the generating function of the cross-cumulant of the real particle momentum and the force-carrier position

$$\Phi \equiv \ln \frac{\tilde{F}(K,q,k,t)}{\tilde{\rho}(K,t)\tilde{W}(q,k,t)} = \ln \frac{\sin(\hbar K \cdot q/2)}{\hbar K \cdot q/2} \qquad (9)$$

As is seen, $\Phi$ is independent of time and of the real particle position. Hence, quantum mechanics is due to disturbances of the virtual particle sea density caused by the motion of the real particle. The only physical parameter in $\Phi$ is the Planck constant $\hbar$ and Eq. (9) can elucidate its physical meaning. The following expansion in power series $\Phi = -(\hbar K \cdot q)^2 / 24 - (\hbar K \cdot q)^4 / 2880 + \cdots$ shows that the second-second cross-cumulant of the virtual particle position and the real particle momentum equals to

$$\kappa_{22} \equiv <(R \cdot p)^2> - <R \otimes R>:<p \otimes p> = -\hbar^2 / 2 \qquad (10)$$

This equation indicates a firm anti-correlation between the virtual particle position and the real particle momentum. In contrast to the famous Heisenberg inequality $\sigma_r \sigma_p \geq \hbar/2$, which restricts the momentum and position of the real quantum particle, the expression from Eq. (10) is an equality, which could be considered as the definition of the Planck constant, $\hbar^2 \equiv -2\kappa_{22}$. A new Heisenberg relation $\sigma_{R^2} \sigma_{p^2} \geq \hbar^2 / 2$ follows from the Cauchy–Schwarz inequality $\kappa_{22} \geq -\sigma_{R^2} \sigma_{p^2}$.